\documentstyle[preprint,aps]{revtex}
\begin{document}

\draft

\title{An Exact Expression of the Collective Excitation Energy Gap of 
Fractional Quantum Hall Effect}

\author{Yu-Liang Liu$^{*}$}
\address{International Center for Theoretical Physics, P. O. Box 586, 34100
 Trieste, Italy}

\maketitle

\begin{abstract}

We have exactly solved the eigenequation of a two-dimensional Dirac 
fermion moving on the surface of a sphere under the influence of a 
radial magnetic field $B$, and obtained an exact expression of the 
collective excitation energy gap for the filling factors 
$\nu=\frac{p}{2mp\pm 1}$, $m$ and $p$ are non-zero integers, which is 
very well agreement with the computing results.

\end{abstract}
\vspace{1cm}


\newpage

Recently, there have been much progress in understanding of the 
fractional quantum Hall effect(FQHE)\cite{1}, which results from a 
strongly correlated incompressible fluid state\cite{2} formed at special 
densities $n_{e}$ of a two-dimensional(2D) electron gas subject to a 
perpendicular magnetic field $B$. 
Based on the idea of "composite fermions", so called by Jain\cite{3}, 
resulting from the highly correlated motion of the carries, which 
effectively attaches even flux quanta to each 
electron[\cite{4}-\cite{8}], Halperin, Lee and Read\cite{9} proposed a 
theory of a compressible Fermi-liquid-like state at the half-filled 
Landau level, $\nu=\frac{1}{2}$, which can be regarded as a degenerate 
system of the composite fermions in the absence of a magnetic field, 
populating a Fermi sea. If the filling factor differs from $\frac{1}{2}$, 
the composite fermions see a net magnetic field $\Delta B=B-B_{1/2}$, 
$B_{1/2}=2n_{e}\Phi_{0}$ ($\Phi_{0}=hc/e$ is the flux quantum), and at 
filling factor $\nu=p/(2p+1)$, and $p$ being a nonzero integer, they may 
fill $|p|$ Landau levels, and correspond to the incompressible quantized 
Hall states whose energy gaps are proportional to the net magnetic fields 
$\Delta B=n_{e}\Phi_{0}/p$. these predictions are strongly supported by 
the current experimental data[\cite{10}-\cite{19}].

On the other hand, there have been a lot of numerical 
calculations[\cite{20}-\cite{25}]employing the spherical 
geometry\cite{26}, in which $N$ electrons move on the surface of a sphere 
under the influence of a radial magnetic field, to study the elementary 
excitations in fractional quantum Hall system. The flux through the 
surface of the sphere is $2S\Phi_{0}$, $S$ is an integer or half-integer, 
the single electron states on the sphere have eigenfunctions $Y_{S,l,m}$, 
called "monopole harmonics", and eigenvalues[\cite{27}, \cite{28}], 
$E_{l,m}=\hbar\frac{2l(l+1)-S^{2}}{2m_{e}R^{2}}$, where $m_{e}$ is the 
electron band mass and $R$ is the radius of the sphere, and satisfies the 
relation $4\pi R^{2}B=2S\Phi_{0}$. The qualitative agreement with the 
numerical results of interaction electrons at the magnetic field $B$ and 
noninteracting fermions at the magnetic field $\Delta B$ gives 
a remarkable confirmation of the composite fermion picture. However, in 
all these approaches, there involves in the electron band mass $m_{e}$ in 
their basic single quasiparticle excitations, but it is widely accepted 
that the FQHE derives from the strongly correlated interaction of the 
electrons within single Landau level, the basic energy unit of the 
quasiparticle excitation is $e^{2}/l_{B}, \; l_{B}=\sqrt{\hbar/(eB)}$ 
being the magnetic length. It is very difficult to scale the electron 
band mass $m_{e}$ by this basic length $l_{B}$\cite{9}. In Ref.\cite{29}, 
we have given a simple explanation for the FQHE at even and odd 
denominator filling factors from the viewpoint of the topologic property 
of a 2D Dirac fermion gas, and used this topologic property to formally 
define a correspondence between the representations of the quantum Hall 
system in terms of electrons and composite fermions in the usual gauge 
field approaches[\cite{4}-\cite{8}]. The basic topologic term of the 2D 
Dirac fermion gas will automatically enforce a Dirac fermion to attach 
with two flux quanta, which becomes a basic composite fermion. In this 
Letter, we give an exact solution of a 2D Dirac fermion moving on the 
surface of a sphere under the influence of a radial magnetic field 
produced by a monopole at the centre of the sphere, and study the 
elementary excitation spectrum in the fractional quantun Hall system.
Of course, the fractional quantum Hall effect derives from the strong
Coulomb interaction of the conduction electrons, to treat it, we must
consider this strong Coulomb interaction of the conduction electrons, however,
up to now we have not found a reasonable method directly to treat it but the
Laughlin's trial wavefunction. We think that the 2D Dirac fermion only
describes a collective excitation mode of this strong Coulomb interaction of
the conduction electrons within single Landau level, therefore there only
exists a Coulomb-like energy scale $e^{2}/l_{B}$ in the 2D Dirac fermion
system, it is a basic energy unit which has not any direct relation with a real
electron mass and the Coulomb interaction of the electrons. For a real
material, its character is reflected only by the dielectric constant
$\epsilon$ which enters the 2D Dirac fermion system with electric charge $e$.
By using the 2D Dirac fermion to study the fractional quantum Hall effect, we
can avoid directly to treat the storngly coupling electronic system.

For a 2D Dirac fermion system, its Hamiltonian can be written in the 
surface of a sphere as\cite{30}
\begin{equation}
H=\int dx \bar{\psi}[-i\gamma^{\mu}D_{\mu}\pm m]\psi
\end{equation}
where $\gamma_{\mu}$ are the 2D Dirac matrices, $\{\gamma_{\mu}, 
\gamma_{\nu}\}=\delta_{\mu\nu}, \; \mu,\nu=1,2$; 
$D_{\mu}=\partial_{\mu}-i(\Gamma_{\mu}+A_{\mu})$, $\Gamma_{\mu}$ are the 
spin-connection, $A_{\mu}$ are the gauge field produced by the monopole 
at the centre of the sphere which is defined as $ 4\pi R^{2}B=2S\Phi_{0}$, 
$R$ is the radius of the sphere, $B$ is an external magnetic field, $2S$ 
(an integer) is the strength of the monopole,
$m$ is the "mass" (chemical potential) of the 2D Dirac fermion which induces a
topologic term (i.e., Chern-Simons term): $\pm\frac{1}{8\pi}\int d\tau d^{2}x
\epsilon_{\mu\nu\lambda}a_{\mu}\partial_{\nu}a_{\lambda}$ in a plane space,
$a_{\mu}$ are the Chern-Simons gauge fields\cite{29'}, this Chern-Simons term
enforces the 2D Dirac fermion attaching with two flux quanta\cite{29},
therefore the 2D Dirac fermion can be seen as an effective composite fermion.
Here we will omit the mass term 
in (1) because it does not affect our final results. Due to in the 2D Dirac 
fermion gas, there naturally exists a topologic term which enforces the 
fermion to carry two flux quanta each\cite{29}, we can use this 2D Dirac 
fermion gas to mimic the strongly correlated electronic system of the 
FQHE, the 2D Dirac fermion should be considered in mind as a collective 
excitation
mode produced by the strong Coulomb interaction of the conduction electrons
within single Landau level, for simplicity, we omit the interaction among the 
2D Dirac 
fermions. To avoid the singularities, we divide the surface of the 
sphere into two regions $R_{a}$ and $R_{b}$, and define two gauge fields 
in these two regions, respectively\cite{27}
\begin{equation}\begin{array}{rl}
A_{r}=A_{\theta}=0,\; & 
A_{\phi}=\displaystyle{\frac{2S}{R\sin\theta}(1-\cos\theta), 
\;\; 0\leq\theta<\frac{\pi}{2}+\delta, \; 0\leq\phi<2\pi}\\
A_{r}=A_{\theta}=0, \; & 
A_{\phi}=\displaystyle{\frac{-2S}{R\sin\theta}(1+\cos\theta),\;\;
\frac{\pi}{2}-\delta<\theta\leq\pi, \; 0\leq\phi<2\pi}
\end{array}\end{equation}
where $\delta$ is a constant, $0<\delta<\frac{\pi}{2}$. From Eq.(1), we 
can obtain the following eigenequation of single 2D Dirac fermion
\begin{equation}
-i\left(\begin{array}{cc}
0,\;\; \frac{A}{R}\\
\frac{\bar{A}}{R},\;\;0
\end{array}\right)\left(\begin{array}{c}
\psi_{1}\\ \psi_{2}
\end{array}\right)=E\left(\begin{array}{c}
\psi_{1}\\ \psi_{2}
\end{array}\right)
\end{equation}
where 
$A=-i(\partial_{\theta}-\frac{i}{\sin\theta}\partial_{\phi}+\frac{\cos\theta}
{2\sin\theta}-RA_{\phi})$, 
$\bar{A}=-i(\partial_{\theta}+\frac{i}{\sin\theta}\partial_{\phi}+
\frac{\cos\theta}{2\sin\theta}+RA_{\phi})$. According to Eq.(3), 
we have the following equations
\begin{equation}\begin{array}{rl}
A\bar{A}\psi_{1}= & R^{2}E^{2}\psi_{1}\\
\bar{A}A\psi_{2}= & R^{2}E^{2}\psi_{2}
\end{array}\end{equation}
Now we define two sets of angular momentum operators
\begin{equation}
\left\{ \begin{array}{ll}
J^{z}= & \mbox{$-i\partial_{\phi}-S^{'}$}\\
J^{\pm}= & \mbox{$\displaystyle{\pm e^{\pm i\phi}\partial_{\theta}+
ie^{\pm i\phi}\frac{\cos\theta}{\sin\theta}\partial_{\phi}
+e^{\pm i\phi}(-\frac{2S-1}{2\sin\theta}+\frac{S^{'}\cos\theta}
{\sin\theta})}$}
\end{array}\right.\end{equation}
\begin{equation}
\left\{ \begin{array}{ll}
L^{z}= &  \mbox{$-i\partial_{\phi}-S^{'}$}\\
L^{\pm}= & \mbox{$\displaystyle{\pm e^{\pm i\phi}\partial_{\theta}+i e^{\pm 
i\phi}\frac{\cos\theta}{\sin\theta}\partial_{\phi}+e^{\pm i\phi}
(-\frac{2S+1}{2\sin\theta}+\frac{S^{'}\cos\theta}{\sin\theta})}$}
\end{array}\right.\end{equation}
where $S^{'}=\pm S$, the total angular momentum operators can be written as
\begin{equation}\begin{array}{rl}
J^{2}= & -\displaystyle{
\frac{1}{\sin\theta}\partial_{\theta}(\sin\theta\partial_{\theta})
-\frac{1}{\sin^{2}\theta}\partial^{2}_{\phi}
-\frac{i}{\sin^{2}\theta}[(2S-1)\cos\theta-2S^{'}]\partial_{\phi}}\\
+ & \displaystyle{
\frac{1}{4\sin^{2}\theta}[(2S-1)\cos\theta-2S^{'}]^{2}+
(S-\frac{1}{2})^{2}}\\
L^{2}= & -\displaystyle{
\frac{1}{\sin\theta}\partial_{\theta}(\sin\theta\partial_{\theta})-
\frac{1}{\sin^{2}\theta}\partial^{2}_{\phi}-\frac{i}{\sin^{2}\theta}
[(2S+1)\cos\theta-2S^{'}]\partial_{\phi}}\\
+ & \displaystyle{
\frac{1}{4\sin^{2}\theta}[(2S+1)\cos\theta-2S^{'}]^{2}+
(S+\frac{1}{2})^{2}}
\end{array}\end{equation}
In the $R_{a}$ region, we have the following relations ($S^{'}=S$)
\begin{equation}\begin{array}{rl}
A\bar{A}= & J^{2}+\displaystyle{\frac{1}{4}}-S^{2}\\
\bar{A}A= & L^{2}+\displaystyle{\frac{1}{4}}-S^{2}
\end{array}\end{equation}
In the $R_{b}$ region, we have the same relations as Eq.(8) if one 
replaces $S^{'}=-S$. The wave functions in these two regions have the 
following relations
\begin{equation}
\psi_{ai}(\theta, \phi)=e^{2iS\phi}\psi_{bi}(\theta, \phi), \;\; i=1,2.
\end{equation}
From Eq.(8), we can easily solve the eigenequation (4). In $R_{a}$ 
region, we have the eigenfunctions
\begin{equation}\begin{array}{rl}
\psi_{1}(\theta,\phi)= & \displaystyle{
e^{i(m+S)\phi}Y^{Sjm}_{1}(\theta), \;\; j=(S-\frac{1}{2}), 
(S-\frac{1}{2})+1, ...,\;\; m=-j, -j+1, ...,j}\\
\psi_{2}(\theta,\phi)= & \displaystyle{
e^{i(n+S)\phi}Y^{Sln}_{2}(\theta), \;\; l=(S+\frac{1}{2}), 
(S+\frac{1}{2})+1, ...,\;\; n=-l,-l+1, ..., l}
\end{array}\end{equation}
where $Y^{Sjm}_{1}(\theta)$ and $Y^{Sln}_{2}(\theta)$ are called the 
monopole harmonics\cite{27}, which satisfy the following equations
\begin{equation}\begin{array}{rl}
 & [j(j+1)-(S-\frac{1}{2})^{2}]Y^{Sjm}_{1}(\theta)=\\ 
& \displaystyle{
[-\frac{1}{\sin\theta}\partial_{\theta}(\sin\theta\partial_{\theta})+
\frac{1}{\sin^{2}\theta}(m+(S-\frac{1}{2})\cos\theta)^{2}]
Y^{Sjm}_{1}(\theta)}  \\
& [l(l+1)-(S+\frac{1}{2})^{2}] Y^{Sln}_{2}(\theta)=\\ 
& \displaystyle{
[-\frac{1}{\sin\theta}\partial_{\theta}(\sin\theta\partial_{\theta})+
\frac{1}{\sin^{2}\theta}(n+(S+\frac{1}{2})\cos\theta)^{2}]
Y^{Sln}_{2}(\theta)}
\end{array}\end{equation}
The eigenvalues of Eq.(4) can be written as
\begin{equation}\begin{array}{rl}
E^{2}_{1}R^{2}= & j(j+1)-\displaystyle{(S-\frac{1}{2})(S+\frac{1}{2}), \;\;
j=(S-\frac{1}{2})+n, \;\; n=0,1,2,...}\\
E^{2}_{2}R^{2}= & l(l+1)-\displaystyle{(S-\frac{1}{2})(S+\frac{1}{2}), \;\;
l=(S+\frac{1}{2})+n, \;\; n=0,1,2,...}
\end{array}\end{equation}
where the quantum numbers $j$ and $l$ satisfy the relation, $l=j+1$. For 
$j=S-\frac{1}{2}$, we only have one set of wave functions
$\psi_{1}(\theta,\phi)=e^{i(m+S)\phi}Y^{S,S-\frac{1}{2},m}_{1}(\theta), 
\; m=-(S-\frac{1}{2}), -(S-\frac{1}{2})+1, ..., S-\frac{1}{2}$, 
corresponding to the eigenvalue $E_{1}(j=S-\frac{1}{2})=0$, called 
"zero-point" energy. For $j=S-\frac{1}{2}+n, \; n$ is a non-zero integer, 
the eigenvalue $E_{1}(j)=E_{2}(l)$ is double degenerate for the 
wavefunctions $\psi_{1}(\theta,\phi)$ and $\psi_{2}(\theta,\phi)$. The 
lowest nonzero eigenvalue of the 2D Dirac fermion is ($j=S+\frac{1}{2}$)
\begin{equation}
E=(1+\frac{1}{2S})^{1/2}\cdot\sqrt{2}E_{B}
\end{equation}
where $E_{B}=\frac{e^{2}}{l_{B}}$ is the Coulomb-like energy of the 2D 
Dirac fermions, $l_{B}$ is the magnetic length. For a free 2D Dirac 
fermion system, $S$ is proportional to the number of the 2D Dirac 
fermions, according to the definition of the filling factor of the 
Landau levels, $\nu=\frac{\rho \Phi_{0}}{B}$, we can easily obtain the 
relation $2S=\frac{N}{\nu}$, so we have the total energy of $N$ 2D Dirac 
fermions from Eq.(13)
\begin{equation}
E_{N}=\sqrt{2} NE_{B}+\frac{\sqrt{2}}{2}\nu E_{B}
\end{equation}
We see that the last term is dependent upon the filling factor $\nu$, 
which can be called as the "collective excitation energy" of the 2D Dirac 
fermion system, $\Delta E_{N}=\frac{\sqrt{2}}{2}\nu E_{B}$. For the 
filling factors $\nu=\frac{p}{2mp\pm 1}$, $p$ and $m$ are nonzero 
integers, the collective excitation energy $\Delta E_{N}$ takes discrete 
values for different $p$ and $m$, and go to the limit $\Delta E_{N}(m, 
p\rightarrow\infty)=\frac{\sqrt{2}}{2}\cdot\frac{1}{2m}E_{B}, \;\; for\; 
p\rightarrow\infty$, which means that there exist infinite nearly 
degenerate levels around these energies for different $m$. It is 
reasonable to assume that the system has a continuous spectrum at these 
energies and show the Fermi-liquid-like behavior\cite{9}. If we choose these 
energies as the orignal point of energy, we can obtain a collective 
excitation energy gap of the system for finite $p$
\begin{equation}
\Delta_{\nu}(m,p)=\mp\frac{\sqrt{2}}{2}\frac{E_{B}}{2m(2mp\pm 1)}
\end{equation}
where the positive and negative signs correspond to the hole and fermion 
filling of the Landau levels.
For a real material we should take $E_{B}=\frac{e^{2}}{\epsilon l_{B}}$, 
$\epsilon$ is the dielectric constant of the material. 
We see that the collective excitation energy gap is proportional to 
$\frac{1}{p}$, but proportional to $\frac{1}{m^{2}}$ for large $m$, this 
behavior is similarly consistent with the experimental data\cite{12}.

We compare the our exact results $\Delta_{\nu}$ with the computing 
results in Ref.\cite{31}. From Eq.(15), we have the following collective 
excitation energy gaps for different filling factors, 
\begin{equation}\begin{array}{rl}
\Delta^{exc.}_{\frac{1}{3}}\simeq 0.118E_{B}, & \;\;
\Delta^{exc.}_{\frac{2}{5}}\simeq 0.071E_{B} \\
\Delta^{exc.}_{\frac{3}{7}}\simeq 0.051E_{B}, & \;\;
\Delta^{exc.}_{\frac{2}{7}}\simeq 0.025E_{B}
\end{array}\end{equation}
while the computing results of Ref.\cite{31} are that, 
$\Delta^{c}_{\frac{1}{3}}\simeq 0.102(3)E_{B}$, 
$\Delta^{c}_{\frac{2}{5}}\simeq 0.063(4)E_{B}$, 
$\Delta^{c}_{\frac{3}{7}}\simeq 0.049(18)E_{B}$,
$\Delta^{c}_{\frac{2}{7}}\simeq 0.022(3)E_{B}$, they are very well 
agreement with our exact results. The small difference between them may 
derive from the computing results extrapoted from finite (a few 
electrons) system. To compare with the experimental data, we must 
redefine Eq.(15). For a fixed $N$(electron number) system, Eq.(15) can be 
changed into as when magnetic field $B$ varies
\begin{equation}
(\frac{E_{N}}{E_{B}})_{\nu=\frac{p}{2mp\pm 
1}}-(\frac{E_{N}}{E_{B}})_{\nu=\frac{1}{2m}}=\mp\frac{\sqrt{2}}{2}\cdot
\frac{1}{2m(2mp\pm 1)}
\end{equation}
This relation is exact for a non-disorder material. However, for an 
integer quantum Hall effect, if the composite fermion 
conception is valid, from Eq.(14) we see that the filling factor $\nu$ 
takes even or odd integer, the system may show different behavior; for 
$\nu=2n$, taking even integer, 
we have the relation, $E_{N}(\nu=2n)=E_{N+n}(\nu=0)$, 
which means that we may have a zero collective excitation energy gap; 
while for $\nu=2n+1$, taking odd integer, we have the relation, 
$E_{N}(\nu=2n+1)=E_{N+n}(\nu=1)$, which means we always have a nonzero 
collective excitation energy gap. If this picture of the integer 
quantum Hall effect is correct, the transitions[\cite{32},\cite{33}] between 
insulating state 
and the quantum Hall states for the filling factors $\nu=1$ and $\nu=2$ 
may be different, because for the former the collective excitation energy 
gap will influence this transition. 

In summary, we have exactly solved the eigenequation of a 2D Dirac 
fermion moving on the surface of a sphere under the influence of a 
radial magnetic field $B$, and obtained an exact expression of the 
collective excitation energy gap for the filling factors 
$\nu=\frac{p}{2mp\pm 1}$, $m$ and $p$ are non-zero integers, which is 
very well agreement with the computing results.

We are very grateful to Prof. L. Yu and Prof. Z. B. Su for their 
encouragement.

*Present address: Max-Planck-Institut f\"{u}r Physik Komplexer Systeme,
Bayreuther Str. 40, D-01187 Dresden, Germany.

\end{document}